\documentclass[preprint,review,12pt]{elsarticle}

\usepackage{amsmath}
\usepackage{amssymb}

\usepackage{graphicx}% Include figure files
\usepackage{bm}% bold math
\renewcommand{\vec}[1]{\boldsymbol{#1}}

\def\cpc{{\itshape Comput. Phys. Commun.} }
\def\pop{{\itshape Phys. Plasmas} }

\journal{Computer Physics Communications}

\begin{document}

\begin{frontmatter}

\title{Multiple Boris integrators for particle-in-cell simulation}

\author[a,b]{Seiji Zenitani\corref{author}}
\author[c]{Tsunehiko N. Kato}

\cortext[author] {Corresponding author.\\\textit{E-mail address:} zenitani@port.kobe-u.ac.jp}

\address[a]{Research Center for Urban Safety and Security, Kobe University, 1-1 Rokkodai-cho, Nada-ku, Kobe 657-8501, Japan}
\address[b]{Research Institute for Sustainable Humanosphere, Kyoto University, Gokasho, Uji, Kyoto 611-0011, Japan}
\address[c]{Center for Computational Astrophysics, National Astronomical Observatory of Japan, 2-21-1 Osawa, Mitaka, Tokyo 181-8588, Japan}

\begin{abstract}
We construct Boris-type schemes for integrating the motion of charged particles in particle-in-cell (PIC) simulation.
The new solvers virtually combine the 2-step Boris procedure arbitrary $n$ times in the Lorentz-force part, and therefore we call them the multiple Boris solvers.
Using Chebyshev polynomials, a one-step form of the new solvers is provided. The new solvers give $n^2$ times smaller errors, allow larger timesteps, and have a long-term stability.
We present numerical tests of the new solvers, in comparison with other particle integrators.
\end{abstract}


\end{frontmatter}

\section{Introduction}

Particle-in-cell (PIC) simulation \citep{birdsall,hockney}
has been extensively used to study complex kinetic phenomena in plasma systems.
The Boris solver \citep{boris70} (or the Boris pusher) is a de fact standard method to advance charged particles in PIC simulation.
Although its algorithm is simple, it is robust and provides moderately good accuracy. For these reasons, it has been used about a half century.

Since mid 2010s, there is a renewed interest to
develop new particle integrators.
In recent PIC simulations, one often solves the system evolution much longer than before.
In such a case, we want to solve the particle motion as accurately as possible, because small errors could be accumulated over many timesteps.
The Boris solver is proven to be stable \citep{qin13,zhang15}, however, it does produce a second-order error. 
For this reason, there is a growing demand for
higher-accuracy extensions to the Boris solver.

\citet{umeda18,umeda19} has proposed multistep Boris solvers.
Repeating a Boris-type procedure multiple steps in the Lorentz-force part,
the author managed to set an effective timestep $2^{N-2}$ times smaller.
Since the Boris solver is a second-order scheme,
$N$-step Boris solvers provide
$4^{N-2}$ times higher accuracy for gyration. 
\citet{zeni18b} have proposed to use a rotation formula in the Lorentz-force part.
The solver virtually provides no error in the phase angle in gyration.
Because of operator-splitting,
their scheme offers second-order accuracy.

In addition to higher numerical accuracy,
it is also important to improve computation cost.
A fast solver helps us to reduce
the total computation time of PIC simulation. 
The cost depends on many other factors, but
the timestep $\Delta t$ is certainly one of them.  
If the solver allows us to use a larger $\Delta t$,
we can further reduce the computation time. 
Thus it is important to evaluate the constraint on $\Delta t$ for particle solvers. 

In this article, we propose
new Boris-type integrators to solve the charged-particle motion. 
The new solvers use a Boris-type procedure multiple times
in the Lorentz-force part.
We call them ``multiple Boris solvers.'' 
They provide higher accuracy and
allow larger timesteps than the original Boris solver.
In Section \ref{sec:boris},
we briefly review the classical 2-step Boris solver
and discuss its graphical meanings.
In Section \ref{sec:multi},
we introduce our multiple Boris solvers.
Using Chebyshev polynomials,
we derive coefficients for $n$-tuple Boris solvers,
where $n$ is an arbitrary integer.
Then we discuss errors, limitations to the timesteps, and
the stability of the phase-space volume. 
In Section \ref{sec:test}, we present benchmark results.
The new solvers and other particle integrators are briefly compared.
Section \ref{sec:discussion} contains discussion and summary.

\section{Boris solver}
\label{sec:boris}

We describe a popular form of the Boris algorithm \citep{birdsall,hockney}.
We refer to it as the classical Boris solver.
The algorithm solves the particle motion in a time-staggered manner,
\begin{align}
\frac{\vec{x}^{t+\Delta t} - \vec{x}^{t}}{\Delta t} &= 
\frac{\vec{u}^{t+\frac{\Delta t}{2}}}
{\gamma^{t+\frac{\Delta t}{2}}}
,
\label{eq:x}
\\
m~\frac{\vec{u}^{t+\frac{\Delta t}{2}} - \vec{u}^{t-\frac{\Delta t}{2}}}{\Delta t} &=
%q \Big( \vec{E}^t + \vec{\bar{v}}^{t} \times \vec{B}^t \Big)
q \bigg( \vec{E}^t + \frac{\vec{u}^{t+\frac{\Delta t}{2}} + \vec{u}^{t-\frac{\Delta t}{2}}}{2\gamma^{t}} \times \vec{B}^t \bigg)
.
\label{eq:u}
\end{align}
The superscripts ($t, t+\Delta t$, ...) indicate time,
$\vec{u}=\gamma\vec{v}$ is the spatial part of a 4-vector, and
$\gamma=[1-(v/c)^2]^{-1/2}=[1+(u/c)^2]^{1/2}$ is the Lorentz factor.
Other symbols have their standard meanings. 
Hereafter we denote $\vec{B}^t$ as $\vec{B}$ for brevity.

The acceleration part (Eq.~\eqref{eq:u}) is
split into
the Coulomb-force part for the first half:
\begin{align}
\vec{u}^{-} &= \vec{u}^{t-\frac{\Delta t}{2}} + \vec{\varepsilon} \Delta t
,
\label{eq:first}
\end{align}
the middle Lorentz-force part:
\begin{subequations}
\begin{align}
\vec{t} &\equiv
\frac{\theta}{2}~ \vec{\hat{b}} = 
\frac{\omega_c \Delta t}{2}~ \vec{\hat{b}} = 
\frac{q\Delta t}{2m \gamma^{-}} \vec{B}
,
\label{eq:boris0} \\
\vec{u}' &=
\vec{u}^{-} + \vec{u}^{-}\times \vec{t}
,
\label{eq:boris1} \\
\vec{u}^{+} &= \vec{u}^{-} + \frac{2}{ 1 + t^2 }
\Big(
\vec{u}'\times \vec{t}
\Big)
,
\label{eq:boris2} 
\end{align}
\end{subequations}
and the Coulomb-force part for the second half:
\begin{align}
\vec{u}^{t+\frac{\Delta t}{2}} &= \vec{u}^{+} + \vec{\varepsilon} \Delta t
.
\label{eq:third}
\end{align}
In these equations, $\vec{\varepsilon} = ({q}/{2m}) \vec{E}^t$,
$\omega_c = ({qB}/{m}\gamma^-) $ is the gyro frequency,
$\theta = 2t = \omega_c\Delta t$ is the gyration angle during $\Delta t$,
$\vec{\hat{b}} = {\vec{B}}/|B|$ is a unit vector in the $\vec{B}$ direction,
$\vec{u}^{-}$ and $\vec{u}^{+}$ are two intermediate states.

\begin{figure}[htbp]
\begin{center}
\includegraphics[width={\columnwidth}]{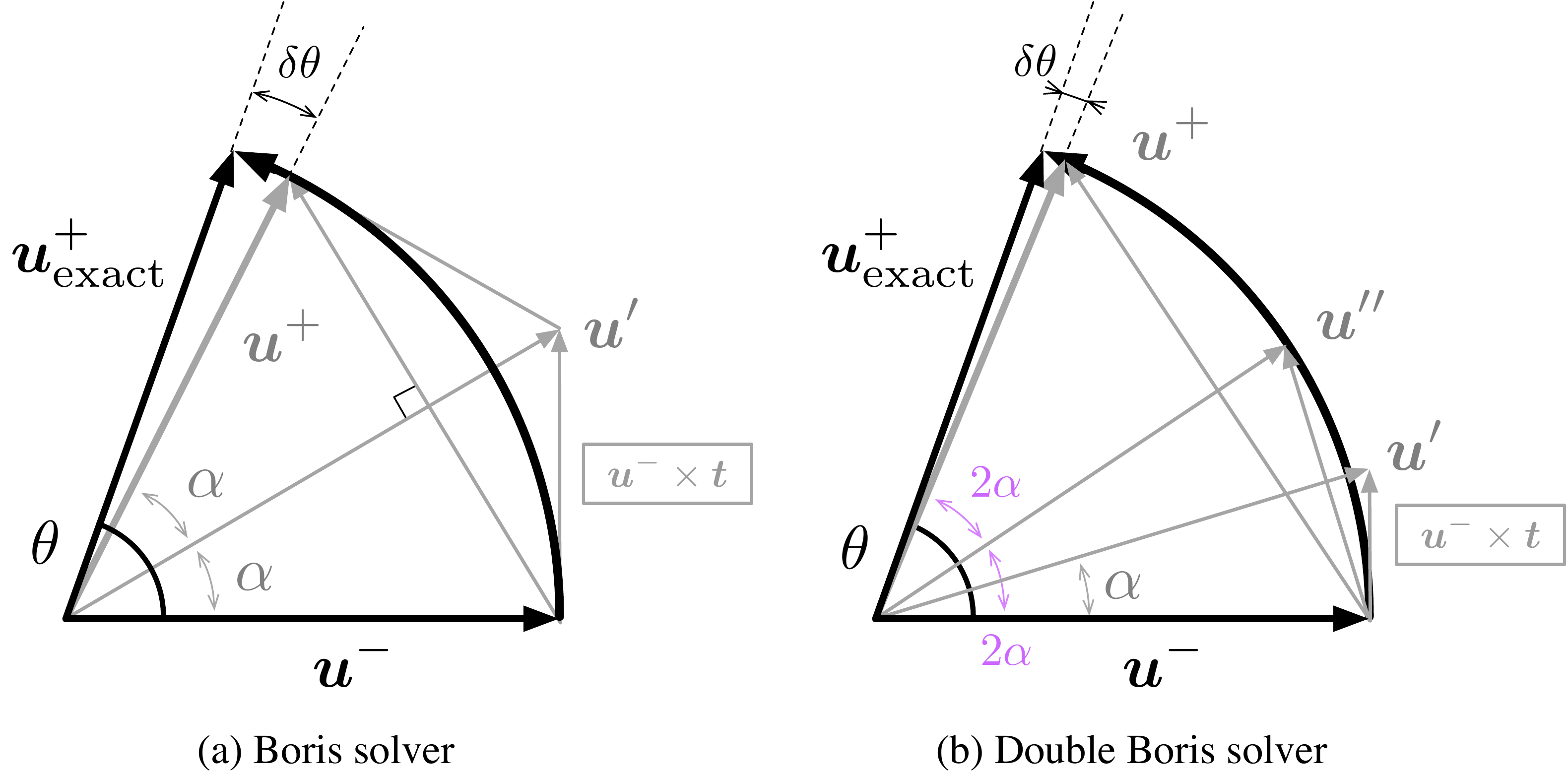}
\caption{
(a) classical Boris solver and
(b) double Boris solver.
\label{fig:illust}}
\end{center}
\end{figure}

We focus on the Lorentz-force part in the middle (Eqs.~\eqref{eq:boris0}--\eqref{eq:boris2}).
This 2-step procedure is illustrated in Figure \ref{fig:illust}(a). 
The coefficient $2/(1+t^2)$ is provided such that
the particle energy remains constant in the Lorentz-force part,
$\gamma^{-}=\gamma^{+}=\gamma^t$.
A base angle $\alpha$ is given by,
\begin{equation}
\alpha \equiv \arctan t
\label{eq:angle}
\end{equation}
We will use the following relations for $\alpha$ in subsequent discussions.
\begin{align}
\sin \alpha = \frac{t}{\sqrt{1+t^2}},~~
\cos \alpha = \frac{1}{\sqrt{1+t^2}},~~
\tan \alpha = t
\label{eq:trig1}
\\
\sin 2\alpha = \frac{2t}{{1+t^2}},~~
\cos 2\alpha = \frac{1-t^2}{{1+t^2}}
\label{eq:trig2}
\end{align}
From Eqs.~\eqref{eq:boris1}, \eqref{eq:boris2}, and \eqref{eq:trig2},
we obtain
\begin{align}
\vec{u}^+
&= \frac{1}{1+t^2}
\bigg( (1-t^2) \vec{u}^-
+
2 ( \vec{u}^- \times \vec{t} )
+
2( \vec{u}^- \cdot \vec{t} ) \vec{t}
\bigg)
\label{eq:borisO}
\\
&=
\vec{u}^- \cos 2\alpha
+
(\vec{u}^- \times \vec{\hat{b}}) \sin 2\alpha
+
(1 - \cos 2\alpha) \vec{u}^-_\parallel
\\
&= 
(\vec{u}^- - \vec{u}^-_\parallel) \cos 2\alpha
+
(\vec{u}^- \times \vec{\hat{b}}) \sin 2\alpha
+
\vec{u}^-_\parallel
,
\label{eq:2a}
\end{align}
where the subscript $\parallel$ indicates the component along $\vec{B}$.
These forms are known as Euler--Rodrigues rotation formula.
The equations correspond to a rotation
about the magnetic field by an approximate angle of $2\alpha$,
instead of the true angle $\theta$.
As indicated in Figure \ref{fig:illust}(a),
the half acceleration in the tangent direction ($\vec{u}^-\times\vec{t}$; Eq.~\eqref{eq:boris1})
corresponds to the half approximate angle $\alpha$
by the Boris solver.
The true rotation angle $\theta$ contains a second-order error of $\delta \theta/\theta \approx -\frac{1}{12} \theta^2$.
\begin{equation}
\theta \approx 2\alpha
=
2 \arctan\frac{\theta}{2}
= \theta \bigg( 1 - \frac{1}{3}\Big(\frac{\theta}{2}\Big)^2 + \frac{1}{5}\Big(\frac{\theta}{2}\Big)^4 - \cdots \bigg)
.\label{eq:error2}
\end{equation}

\section{Multiple Boris solver}
\label{sec:multi}

\citet{umeda18} has proposed a novel way to
reduce the error ($\delta \theta/\theta$).
He employed a half timestep $\Delta t \rightarrow \frac{1}{2}\Delta t$ in the Lorentz-force part (Eqs.~\eqref{eq:boris0}--\eqref{eq:boris2})
and then repeated the Boris procedure twice.
Importantly, the timestep for the position (Eq.~\eqref{eq:x}) and the Coulomb-force parts (Eqs.~\eqref{eq:first} and \eqref{eq:third}) are unchanged.
We call this method the double Boris solver, and
it is illustrated in Figure \ref{fig:illust}(b).
Instead of Eq.~\eqref{eq:boris0}, we use
\begin{align}
\vec{t} \equiv
\frac{\theta}{4}~ \vec{\hat{b}}
=\frac{q\Delta t}{4m \gamma^-} \vec{B}
,~~~
\alpha \equiv \arctan t
\label{eq:4t}
\end{align}
The base angle $\alpha$ is redefined,
as indicated in Figure \ref{fig:illust}(b).
Then, we repeat the 2-step Boris procedure
(Eqs.~\eqref{eq:boris1} and \eqref{eq:boris2}) twice.
We obtain a half state $\vec{u}''$ by the first Boris procedure,
which is given by Eq.~\eqref{eq:2a} with Eq.~\eqref{eq:4t}.
Applying $\cdot\vec{\hat{b}}$ to Eq.~\eqref{eq:2a},
one can confirm that the parallel component is preserved,
$\vec{u}''_{\parallel}=(\vec{u}''\cdot\vec{\hat{b}})\vec{\hat{b}} = \vec{u}^-_\parallel$.
Substituting $\vec{u}''$ to Eq.~\eqref{eq:2a} again,
we obtain the next state $\vec{u}^{+}$,
\begin{align}
\vec{u}^{+}
&= 
(\vec{u}'' - \vec{u}^-_\parallel) \cos 2\alpha
+
(\vec{u}'' \times \vec{\hat{b}}) \sin 2\alpha
+
\vec{u}''_\parallel \\
&=
( \vec{u}^- - \vec{u}^-_\parallel )
\cos 4\alpha
+
(\vec{u}^- \times \vec{\hat{b}})
\sin 4\alpha
+
\vec{u}^-_\parallel
\end{align}
The second Boris operation advances the phase angle by another $2\alpha$, and
the entire procedure gives gyration by $4\alpha$.
In Ref.~\cite{umeda18}, the author proposed a 3-step procedure
to repeat the Boris operation twice.
Now we obtain a smaller second-order error in gyration,
$\delta \theta/\theta \approx -\frac{1}{48} \theta^2$.
\begin{equation}
\theta \approx 4\alpha
=
4 \arctan\frac{\theta}{4}
= \theta \bigg( 1 - \frac{1}{3}\Big(\frac{\theta}{4}\Big)^2 + \frac{1}{5}\big(\frac{\theta}{4}\Big)^4 - \cdots \bigg)
.\label{eq:error4}
\end{equation}

Let us extend this idea.
In order to better approximate the gyro motion,
we propose to repeat the Boris procedure $n$ times,
where $n$ is an arbitrary positive integer.
We call the resulting schemes ``multiple Boris solvers.''
Here below, we describe its procedure in the Lorentz-force part.

We define the tangent vector $\vec{t}$ and
the base angle $\alpha$ as follows.
\begin{equation}
\vec{t} \equiv \frac{\theta}{2n}\vec{\hat{b}}
=\frac{q\Delta t}{2nm \gamma^-} \vec{B}
,~~~
\alpha \equiv \arctan t
.
\label{eq:angle_n}
\end{equation}
Note that $\vec{t}$ is divided by $n$.
From the discussion on the double Boris solver,
repeating the Boris operation $n$ times,
one can obtain a rotation angle of $2n\alpha$.
The next state $\vec{u}^+_n$ should be given by
\begin{equation}
\vec{u}^+_n
= 
(\vec{u}^{-}-\vec{u}^-_\parallel)\cos(2n\alpha) 
+
( \vec{u}^{-} \times \vec{\hat{b}} )
\sin(2n\alpha)
+
\vec{u}^-_\parallel
\label{eq:ma}
\end{equation}
The subscript $n$ indicates quantities for the $n$-tuple solver. 
For brevity, we hereafter omit the minus sign from $\vec{u}^{-}$.
Since we desire a simulation-friendly expression
with $\vec{t}$,
we modify Eq.~\eqref{eq:ma} to
\begin{align}
\vec{u}^+_n
&=
\cos(2n\alpha) \vec{u}
+
\frac{\sin(2n\alpha)}{t} ( \vec{u} \times \vec{t} )
+
\frac{ 2\sin^2(n\alpha) }{t^2}
(\vec{u}\cdot\vec{t})\vec{t}
\label{eq:n_angle}
\\
&= 
c_{n1} \vec{u}
+
c_{n2} ~ (\vec{u} \times \vec{t})
+
c_{n3}
( \vec{u} \cdot \vec{t} ) \vec{t}
\label{eq:ut}
\end{align}
where $c_{n1}$, $c_{n2}$, and $c_{n3}$ are coefficients.

Next, we construct generic expressions for these coefficients.
We use Chebyshev polynomials of the first kind $T_n(x): T_0(x) = 1, ~T_1(x) = x, ~T_2(x) = 2x^2-1, \dots$ and the second kind $U_n(x): U_0(x) = 1, ~U_1(x) = 2x, ~U_2(x) = 4x^2-1, \dots$.
They satisfy useful relations:
\begin{equation}
\cos n x = T_n (\cos x),~~
\sin n x = \sin x ~ U_{n-1} (\cos x)
\label{eq:chebyshev}
\end{equation}
We also employ
\begin{equation}
p \equiv \cos 2\alpha = \frac{1-t^2}{1+t^2},
~~~
1+p = \frac{\sin 2\alpha}{t} = \frac{2}{1+t^2}
.
\label{eq:p}
\end{equation}
With help from Eqs.~\eqref{eq:trig1}, \eqref{eq:trig2}, \eqref{eq:chebyshev}, and \eqref{eq:p},
we express the coefficients in Eq.~\eqref{eq:ut} in the following way,
\begin{align}
c_{n1}
&=
T_{n}(\cos 2\alpha)
=
T_{n}(p)
\label{eq:p1}
\\
c_{n2}
&=
\frac{1}{t} \sin 2\alpha~
U_{n-1} (\cos 2\alpha)
=
(1+p) U_{n-1}(p)
\label{eq:p2}
\\
c_{n3}
&=
\left\{
\begin{array}{ll}
{1+p}
& {\rm (for~{\it n}=1)}
\\
{(1+p)}~
\Big(
U_{k}(p)
+
U_{k-1}(p)
\Big)^2
& {\rm (for~{\it n}=2k+1)}
\\
%\frac{2}{t^2}
%\Big(
%\sin 2\alpha
%~
%U_{k-1}
%(\cos 2\alpha)
%\Big)^2
%& \\
%~~~=
2~
\bigg(
(1+p)
U_{k-1}(p)
\bigg)^2
%& {\rm (for~even~{\it n})}
& {\rm (for~{\it n}=2k)}
\end{array}
\right.
\label{eq:p3}
\end{align}
%%% One can test the p-formula at Wolfram Alpha https://www.wolframalpha.com/.
%%% ChebyshevT[2, x]
%%% ChebyshevU[2-1, x] * (1+x)
%%% ( ChebyshevU[4/2-1, x] * (1+x) )^2 * 2
Here, $k$ is a positive integer.
We derive Eq.~\eqref{eq:p3} for $n=2k+1$ ($n=3,5,\dots$) in \ref{sec:2k+1}. 
%Note that $c_{n1}=1-c_{n3}t^2$.
Practically, we compute $p$ first, and then
calculate Eqs.~\eqref{eq:p1}-\eqref{eq:p3} by using $p$.
Some higher-order polynomials can be calculated
via nesting relations \citep{rivin90}:
\begin{equation}
T_{mn}(x)=T_m(T_n(x)),~
U_{mn-1}(x) = U_{m-1}(T_n(x))U_{n-1}(x)
\end{equation}
%For example, $T_{2n}(x)=2(T_n(x))^2-1$ and $U_{2n-1}(x) = 2T_n(x)U_{n-1}(x)$ are useful in the $n=2^N$ cases.

We can also express the coefficients with $t$.
From the polynomials,
\begin{align}
c_{n1}
&=
%%\cos(2n\alpha) = 
%T_{n}(\cos 2\alpha)
%=
T_{n} \Big(\frac{1-t^2}{1+t^2}\Big)
\label{eq:t1}
\\
c_{n2}
&=
%%\frac{1}{t} \sin(2n\alpha) =
%\frac{1}{t} \sin 2\alpha~
%U_{n-1} (\cos 2\alpha)
%=
\frac{2}{1+t^2}
U_{n-1} \Big(\frac{1-t^2}{1+t^2}\Big)
\label{eq:t2}
\\
c_{n3}
%%&= \frac{1}{t^2} \big[1-\cos(2n\alpha)\big]
%%\nonumber \\
&=
\left\{
\begin{array}{ll}
%%\frac{1}{t^2}
%%\Big(
%%1-
%%T_{n}\big(\frac{1-t^2}{1+t^2}\big)
%%\Big)
%%& {\rm (for~odd~{\it n})}
%%\\
%%\frac{2}{t^2}
%%\Big(
%%\sin \alpha
%%~
%%U_{n-1}
%%(\cos \alpha)
%%\Big)^2
%%& \\
%%~~~=
%%\frac{2}{(1+t^2)}
%%\Big(
%%U_{{n}-1}
%%\big(\frac{1}{\sqrt{1+t^2}}\big)
%%\Big)^2
%%& {\rm (for~odd~{\it n})}
%%\\
\frac{2}{1+t^2}
& {\rm (for~{\it n}=1)}
\\
\frac{2}{1+t^2}
\Big(
U_{k}\big(\frac{1-t^2}{1+t^2}\big)
+
U_{k-1}\big(\frac{1-t^2}{1+t^2}\big)
\Big)^2
& {\rm (for~{\it n}=2k+1)}
%%{\rm (for~odd~{\it n})}
\\
%\frac{2}{t^2}
%\Big(
%\sin 2\alpha
%~
%U_{k-1}
%(\cos 2\alpha)
%\Big)^2
%& \\
%~~~=
\frac{8}{(1+t^2)^2}
\Big(
U_{k-1}
\big(\frac{1-t^2}{1+t^2}\big)
\Big)^2
& {\rm (for~{\it n}=2k)}
%%& {\rm (for~even~{\it n})}
\end{array}
\right.
\label{eq:t3}
\end{align}
%%% One can test the t-formula by Wolfram Alpha https://www.wolframalpha.com/
%%% ChebyshevT[2, (1-t^2)/(1+t^2)]
%%% ChebyshevU[2-1, (1-t^2)/(1+t^2)] * 2 / (1+t^2)
%%% ( ChebyshevU[n-1, 1/ Sqrt[1+t^2]] )^2 * 2 / (1+t^2)
%%% ( ChebyshevU[4/2-1, (1-t^2)/(1+t^2)] )^2 * 8 / (1+t^2)^2
we obtain explicit forms, for example,
\begin{align}
c_{11} = \frac{1-t^2}{1+t^2},~~
c_{12} = \frac{2}{1+t^2},~~
c_{13} = \frac{2}{1+t^2},
\label{eq:c1}
\\
c_{21} = \frac{1-6t^2+t^4}{(1+t^2)^2},~~
c_{22} = \frac{4(1-t^2)}{(1+t^2)^2},~~
c_{23} = \frac{8}{(1+t^2)^2},
\label{eq:c2}
\end{align}
\begin{align}
c_{31} = \frac{1-15t^2+15t^4-t^6}{(1+t^2)^3},~~\nonumber\\
c_{32} = 
%\frac{2 (9 - 6 t^2 + t^4))}{(1+t^2)^3},~~
\frac{2 (t^2-3)(3t^2-1)}{(1+t^2)^3},~~
c_{33} = \frac{2(t^2-3)^2}{(1+t^2)^3},
\label{eq:c3}
\\
c_{41}
= 
\frac{\big((1-t^2)^2 - 24t^2\big)(1-t^2)^2 + 16t^4}{(1+t^2)^4}~~\nonumber\\
= \frac{1-28t^2+70t^4-28t^6+t^8}{(1+t^2)^4},~~\nonumber\\
c_{42} = \frac{8(1-t^2)(1-6t^2+t^4)}{(1+t^2)^4},~~
%c_{42} = \frac{8(1-7t^2+7t^4-t^6)}{(1+t^2)^4},~~
c_{43} = \frac{32(1-t^2)^2}{(1+t^2)^4}
\label{eq:c4}
\end{align}
It is interesting to see that the $c_{n1}$, $c_{n2}$, and $c_{n3}$ share the same denominator, $(1+t^2)^n$.
One can confirm this by inspecting Eqs.~\eqref{eq:t1}--\eqref{eq:t3}.

As a result, the multiple Boris solver advances the velocity
by using Eqs.~\eqref{eq:first}, \eqref{eq:angle_n}, \eqref{eq:p}--\eqref{eq:p3}, \eqref{eq:ut}, and \eqref{eq:third}.
One can also use the coefficients in Eqs.~\eqref{eq:c1}--\eqref{eq:c4}
instead of Eqs.~\eqref{eq:p}--\eqref{eq:p3}.

\begin{figure}[htbp]
\begin{center}
\includegraphics[width={0.9\columnwidth}]{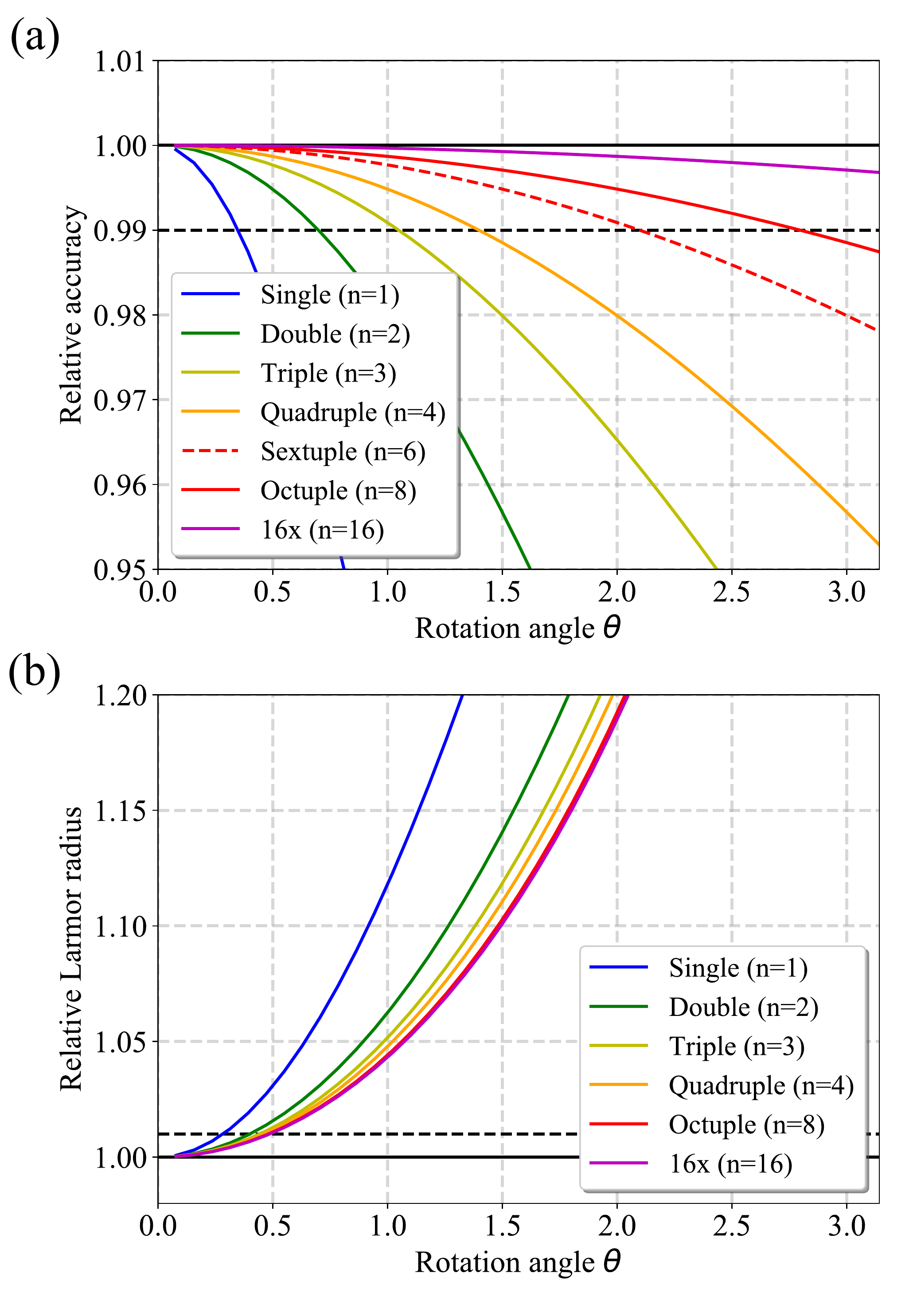}
\caption{
(a) Relative accuracy in phase angle and
(b) relative length of Larmor radius
for the multiple Boris solvers.
\label{fig:curve}}
\end{center}
\end{figure}

The Lorentz-force part of the multiple Boris solvers provides
a second-order error in phase angle.
From Eq.~\eqref{eq:angle_n}, we obtain
\begin{equation}
\theta \approx 2n\alpha
=
2n\arctan\frac{\theta}{2n}
= \theta~\bigg( 1 - \frac{\theta^2}{12 n^2} + \frac{\theta^4}{80 n^4} - \cdots \bigg)
.\label{eq:error}
\end{equation}
Thus, the multiple Boris solvers
provide $n^2$ times smaller errors than the single Boris solver.
Eq.~\eqref{eq:error} approaches the exact angle for $n\rightarrow \infty$. 

In Figure \ref{fig:curve}(a),
we plot relative accuracy in phase angle, $\theta^{-1}\cdot2n\arctan({\theta}/{2n})$,
for several $n$.
In the classical Boris solver,
which is equivalent to the $n=1$ case,
we typically set $\theta = \omega_c\Delta t < 0.2$--$0.3$ to keep an error within 1\%.
Considering ${\theta^2}/({12 n^2}) \lesssim 0.01$, we obtain
\begin{align}
\theta \lesssim 0.35 n
.
\label{eq:deltat_1}
\end{align}
The multiple Boris solvers allow
$n$-times larger timestep than the single Boris solver. 

Next, we evaluate an effective Larmor radius.
We consider a simple gyration about $\vec{B}$
in the case of $\vec{E}=0$.
From Eq. \eqref{eq:x}, we obtain
\begin{align}
\big|\vec{x}^{t+\Delta t} - \vec{x}^{t}\big|
= \frac{\big| \vec{u}^{t+\frac{\Delta t}{2}} \big| }
{\gamma^{t+\frac{\Delta t}{2}}}\Delta t
= |v_R| \Delta t
,
\label{eq:radius1}
\end{align}
where $v_R$ is the velocity.
The two points $\vec{x}^{t+\Delta t}$ and $\vec{x}^{t}$ are located on
the same circle with a Larmor radius of $r_L'$.
The two points should be two endpoints of
an arc with a central angle $2n\alpha$.
Then we expect
\begin{align}
|\vec{x}^{t+\Delta t} - \vec{x}^{t}|
= 2r_L' \sin \Big(\frac{2n\alpha}{2}\Big)
.
\label{eq:radius2}
\end{align}
From Eqs.~\eqref{eq:radius1} and \eqref{eq:radius2},
we obtain
\begin{align}
r_L'
= \frac{|v_R| \Delta t}{2\sin(n\alpha)}
= r_L~\frac{\theta}{2\sin(n\alpha)}
\label{eq:radius3}
\end{align}
where $r_L=|v_R|/\omega_c$ is a physical Larmor radius.
For $n=1$,
we obtain a familiar result of $r'_L = r_L [1+(\theta/2)^2]^{1/2}$
from Eq.~\eqref{eq:trig1}.
In general cases, Eq.~\eqref{eq:radius3} can be expanded to
\begin{align}
r_L'
= r_L~\bigg( 1 + \Big[ \frac{1}{24} + \frac{1}{12n^2} \Big] \theta^2
+ \frac{7n^4-20n^2-32}{5760n^4} \theta^4
 + \mathcal{O}(\theta^6) \bigg)
\label{eq:radius4}
\end{align}

Figure \ref{fig:curve}(b) shows
the relative length of Larmor radius, $r'_L/r_L = {\theta}/[2\sin(n\alpha)]$.
In the square brackets in Eq.~\eqref{eq:radius4},
the first term is insensitive to $n$, while
the second depends on $n$.
Thus, higher $n$ does not allow $n$-times larger timestep.
To keep an error within $\approx 1\%$,
we need to limit the timestep to
\begin{align}
\theta = \omega_c\Delta t \lesssim 0.5 \sqrt{\frac{n^2}{n^2+2}}
.
\label{eq:deltat_2}
\end{align}
For $n\rightarrow \infty$, we find $\omega_c\Delta t < 0.5$.
If we set the threshold to $\epsilon\%$, 
both Eqs.~\eqref{eq:deltat_1} and \eqref{eq:deltat_2} can be relaxed by a factor of $\sqrt{\epsilon}$.
For example, Eq. \eqref{eq:deltat_2} gives $\omega_c\Delta t < 1$ for $4\%$.

The multiple Boris solvers have two nice properties for stable solvers.
The first one is the time symmetry.
As evident in Eqs.~\eqref{eq:first}, \eqref{eq:ma}, and \eqref{eq:third},
the multiple Boris solvers is symmetric in time. 
In such a case, it is unlikely that
the solvers numerically damp or amplify a motion in one time direction.
In Eq.~\eqref{eq:ut}, procedures in the reverse-time direction are obtained
by reversing the sign of the second coefficient,
i.e., $c_{-n1} = c_{n1}, c_{-n2} = -c_{n2}$, and $c_{-n3} = c_{n3}$.
It is obvious that $c_{01} = 1, c_{02} = 0$, and $c_{03} = 0$ for $n=0$.
As a result, we have the coefficients
not only for positive integers, but also for any integer $n$.

The second is the volume preservation \citep{qin13,zhang15},
which is related to the long-term stability. 
We consider the evolution of a volume in the phase-space
from $t-\frac{\Delta t}{2}$ to ${t+\frac{\Delta t}{2}}$. 
Since the Lorentz-force part of the multiple Boris solver
is equivalent to $n$-times repetition of the Boris-type element,
we express the equivalent numerical procedures
into the following steps \citep{zhang15} of $S_1, S_2, \cdots S_5$,
where $S_k$ indicates the $k$-th step.
The third step, the Lorentz-force part of the solver,
is further divided into $n$ substeps.
{\tt Boris}$(\vec{u}, \Delta t)$ indicates the Boris-type procedure (Eqs.~\eqref{eq:boris0}--\eqref{eq:boris2}) on the vector $\vec{u}$ for a timestep $\Delta t$.
\begin{align}
&S_1: &
\vec{x}^{t} &=
\vec{x}^{t-\frac{\Delta t}{2}} + \frac{\Delta t}{2} \frac{\vec{u}^{t-\frac{\Delta t}{2}}}{\gamma^{t-\frac{\Delta t}{2}}},
&
\vec{u}^{t-\frac{\Delta t}{2}} &= \vec{u}^{t-\frac{\Delta t}{2}}
\label{eq:VPA1}\\
&S_2: &
\vec{x}^{t} &= \vec{x}^{t},
&
\vec{u}^{-} &= \vec{u}^{t-\frac{\Delta t}{2}} + \vec{\varepsilon} \Delta t
\label{eq:VPA2}\\
&S_{31}: &
\vec{x}^{t} &= \vec{x}^{t},
&
\vec{u}^{+}_1 &= \mathtt{Boris}( \vec{u}^{-}, \frac{\Delta t}{n} )
\label{eq:VPA31}\\
&&&& \vdots \nonumber \\
&S_{3m}: &
\vec{x}^{t} &= \vec{x}^{t},
&
\vec{u}^{+}_m &= \mathtt{Boris}( \vec{u}^{+}_{m-1}, \frac{\Delta t}{n} )
\label{eq:VPA3m}\\
&&&& \vdots \nonumber \\
&S_{3n}: &
\vec{x}^{t} &= \vec{x}^{t},
&
\vec{u}^{+}_n &= \mathtt{Boris}( \vec{u}^{+}_{n-1}, \frac{\Delta t}{n} )
\label{eq:VPA3n}\\
&S_{4}: &
\vec{x}^{t} &= \vec{x}^{t},
&
\vec{u}^{t+\frac{\Delta t}{2}} &= \vec{u}^{+}_n + \vec{\varepsilon} \Delta t
\label{eq:VPA4}\\
&S_{5}: &
\vec{x}^{t+\frac{\Delta t}{2}} &=
\vec{x}^{t} + \frac{\Delta t}{2} \frac{\vec{u}^{t+\frac{\Delta t}{2}}}{\gamma^{t+\frac{\Delta t}{2}}},
&
\vec{u}^{t+\frac{\Delta t}{2}} &= \vec{u}^{t+\frac{\Delta t}{2}}
\label{eq:VPA5}
\end{align}
We check the Jacobian $J_k$ for the $k$-th step, $S_k$.
\begin{align}
J_k
=
\Big| \frac{\partial(\vec{x}^{new},\vec{u}^{new})}{\partial(\vec{x}^{old},\vec{u}^{old})} \Big|
=
{\rm det}
\begin{pmatrix}
\cfrac{\partial \vec{x}^{new}}{\partial \vec{x}^{old}} &
\cfrac{\partial \vec{x}^{new}}{\partial \vec{u}^{old}} \\
\cfrac{\partial \vec{u}^{new}}{\partial \vec{x}^{old}} &
\cfrac{\partial \vec{u}^{new}}{\partial \vec{u}^{old}} \\
\end{pmatrix}
.
\end{align}
For the first two and the last two steps ($S_1$, $S_2$, $S_4$, and $S_5$), we obtain
\begin{align}
J_1 = J_5 =
\begin{vmatrix}
{I} &
\frac{\partial \vec{x}^{new}}{\partial \vec{u}^{old}} \\
0 & {I} \\
\end{vmatrix}
= 1,
~~
J_2 = J_4 =
\begin{vmatrix}
{I} & 0 \\
\frac{\partial \vec{u}^{new}}{\partial \vec{x}^{old}}
& {I} \\
\end{vmatrix}
= 1
\end{align}
In addition, \citet{zhang15} proved that
the Lorentz-force part of the relativistic Boris solver ({\tt Boris}$(\vec{u}, {\Delta t})$)
preserves a phase-space volume,\footnote{The authors claimed that they developed a new volume-preserving algorithm, but the algorithm appears to be the standard relativistic Boris solver.} and
the proof holds true for an arbitrary timestep $\Delta t$. 
Then, for the substeps in the third step ($S_{31}, \cdots S_{3m}, \cdots S_{3n}$),
each Boris-type element ({\tt Boris}$(\vec{u}, \frac{\Delta t}{n})$) preserves the phase-space volume, i.e., $J_{31} = \cdots = J_{3m} = \cdots = J_{3n} = 1$. 
As a result, the Jacobian throughout the entire procedure satisfies
$J_{\rm all} = J_1 J_2 ( J_{31} \cdots J_{3m} \cdots J_{3n} ) J_4 J_5 = 1$.
This indicates that the entire procedure preserves the phase-space volume. 
The multiple Boris solvers are volume-preserving.

\section{Numerical tests}
\label{sec:test}

To evaluate the new solvers,
we have conducted four numerical tests.
For comparison,
the following ten particle integrators are benchmarked:
\begin{enumerate}
\item
Classical Boris solver
(Eqs.~\eqref{eq:boris0}--\eqref{eq:boris2}),
\item
Single Boris solver ($n=1$; Eqs.~\eqref{eq:ut}, \eqref{eq:c1})
\item
Double Boris solver ($n=2$; Eqs.~\eqref{eq:ut}, \eqref{eq:c2})
\item
Quadruple Boris solver ($n=4$; Eqs.~\eqref{eq:ut}, \eqref{eq:c4}).
\item
Octuple Boris solver ($n=8$; Eqs.~\eqref{eq:ut}, \eqref{eq:p}--\eqref{eq:p3}).
\item
$16$x Boris solver ($n=16$; Eqs.~\eqref{eq:ut}, \eqref{eq:p}--\eqref{eq:p3}).
\item
$32$x Boris solver ($n=32$; Eqs.~\eqref{eq:ut}, \eqref{eq:p}--\eqref{eq:p3}).
\item
Umeda solver ($N=3$) \citep{umeda18}
\item
Umeda solver ($N=4$) \citep{umeda19}
\item
Exact-gyration solver \citep{zeni18b}
\end{enumerate}

The first one is the classical Boris solver \citep{birdsall,hockney}.
The next six are the multiple Boris solvers
with $n=1$, $n=2$, $n=4$, $n=8$, $n=16$, and $n=32$.
The coefficients are given by
Eqs.~\eqref{eq:c1}--\eqref{eq:c4} or Eqs.~\eqref{eq:p}--\eqref{eq:p3}. 
We refer to them as the single, double, quadruple, octuple, $16$x and $32$x Boris solvers, respectively.
The next two are the Umeda solvers with $N=3$ steps \citep{umeda18} and $N=4$ steps \citep{umeda19}.
We follow procedures in Eqs.~(8)--(10) in Ref.~\citep{umeda19}.
As they contain $2^{N-2}$ Boris elements,
they are equivalent to the double and quadruple Boris solvers.
The last one is \citet{zeni18b}'s solver,
which directly calculates Euler's rotation formula
in the Lorentz-force part:
\begin{align}
\label{eq:new1}
\theta &\equiv \frac{qB}{m \gamma} \Delta t, \\
\vec{u}_{\parallel} &= (\vec{u}\cdot\vec{\hat{b}})~\vec{\hat{b}} \\
\label{eq:new2}
\vec{u}^{+} &= \vec{u}_{\parallel} + (\vec{u}-\vec{u}_{\parallel}) \cos \theta + ({\vec{u} \times \vec{\hat{b}}}) \sin \theta
\end{align}
By definition, this solver does not produce an error during the gyration phase.
We refer to it as the exact-gyration solver.
Since it uses trigonometric functions, its computation cost is usually higher than those of polynomial-based solvers. 
Note that all these solvers share the same Coulomb-force procedures (Eqs:~\eqref{eq:first} and \eqref{eq:third}).

\begin{figure}[htbp]
\begin{center}
\includegraphics[width={\columnwidth}]{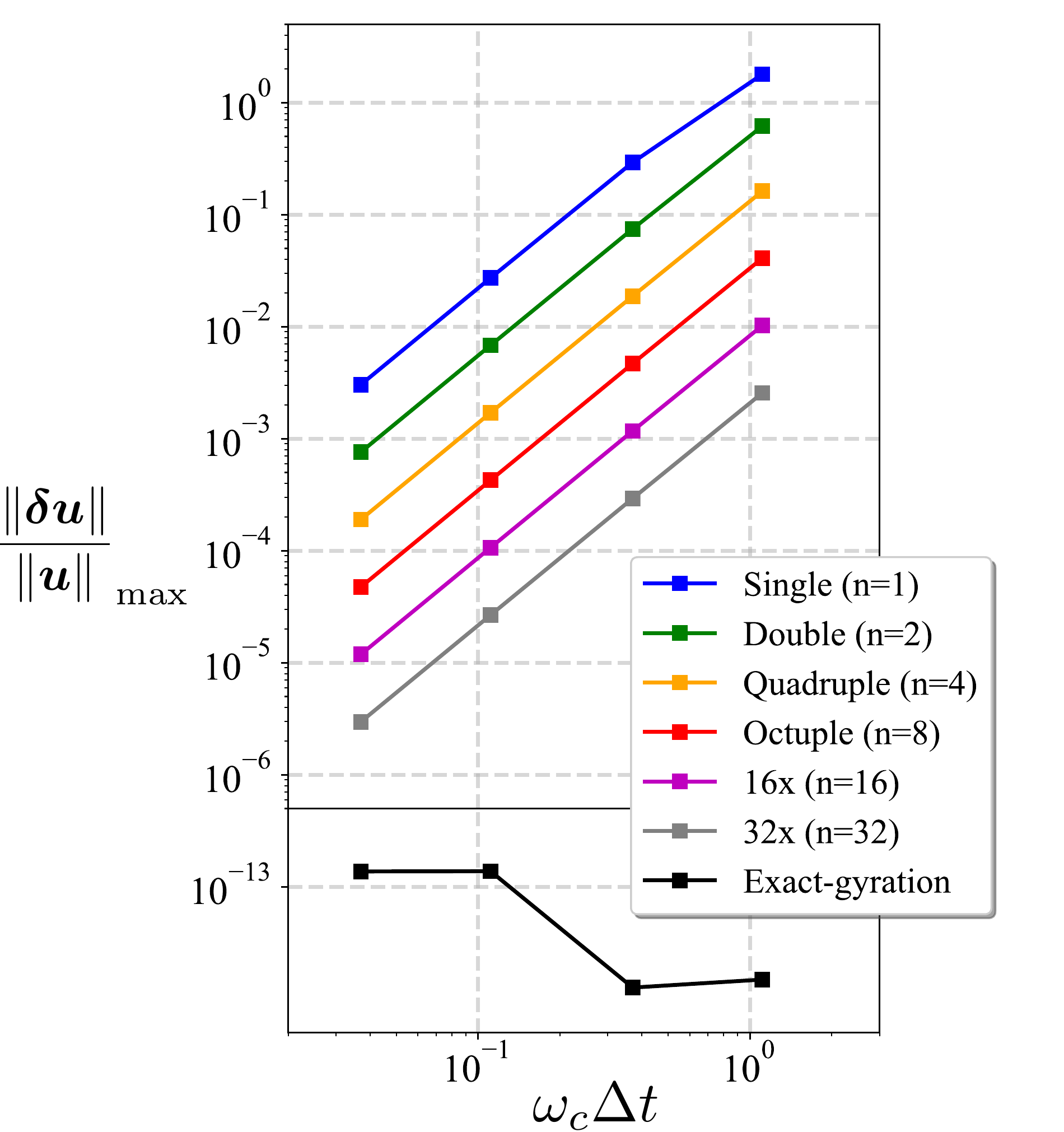}
\caption{
Normalized errors in $\vec{u}$ by the test-particle simulations as a function of $\theta=\omega_c\Delta t$.
\label{fig:scheme}}
\end{center}
\end{figure}

In the first test,
we have run test-particle simulations in a static magnetic field $\vec{B}=(0,0,1)$ and $\vec{E}=(0,0,0)$.
Parameters are set to $m=1$, $q=1$, and $c=1$.
The initial condition is $\vec{u}(t=0)=(1,0,0)$ and the timestep is set to $\Delta t = \pi/60$, $\pi/20$, $\pi/6$, or $\pi/2$.
We have calculated errors in $\vec{u}$ from the analytic solutions.
We have run the simulations until $t=12\pi$, and then
we have evaluated the maximum errors in a 4-vector,
$\{|\delta\vec{u}|/|\vec{u}|\}_{\rm max}$. 
Since $\omega_c=qB/(m\gamma)=1/\sqrt{2}$,
the interval $12\pi$ corresponds to $\approx 4$ rotations.
This test is a subset of numerical tests in our recent study \citep{zeni18b}. 
Here, almost all of the solvers are combinations of
the exact solver for the Coulomb-force parts (Eqs.~\eqref{eq:first} and \eqref{eq:third}) and
the second-order solver for the Lorentz-force part.
These solvers only differ in the Lorentz-force part. 
%Better Lorentz-force solvers usually provide better results
%in general cases of $\vec{B}\ne 0$ and $\vec{E}\ne 0$.

Figure \ref{fig:scheme} shows the results
as a function of $\Delta t$.
Errors by the classical Boris solver and the Umeda solvers are not presented, because the errors are identical to those by the single, double, and quadruple Boris solvers.
One can see that
the multiple Boris schemes give second-order errors with respect to $\Delta t$.
The errors also scale like $\propto n^{-2}$.
This is because 
the error simply accumulates
in the interval of our interest.
From Eq.~\eqref{eq:error},
the errors can be estimated to
$\delta\theta \approx (\omega_c 12\pi)\frac{1}{12}(\omega_c\Delta t/n)^2 = (\pi / \sqrt{2})(\omega_c\Delta t/n)^2$.
The results are in excellent agreement with the prediction.
Meanwhile, the exact-gyration solver only gives small errors of $\mathcal{O}(10^{-13})\sim\mathcal{O}(10^{-14})$. 

\begin{figure}[htbp]
\begin{center}
\includegraphics[width={\columnwidth}]{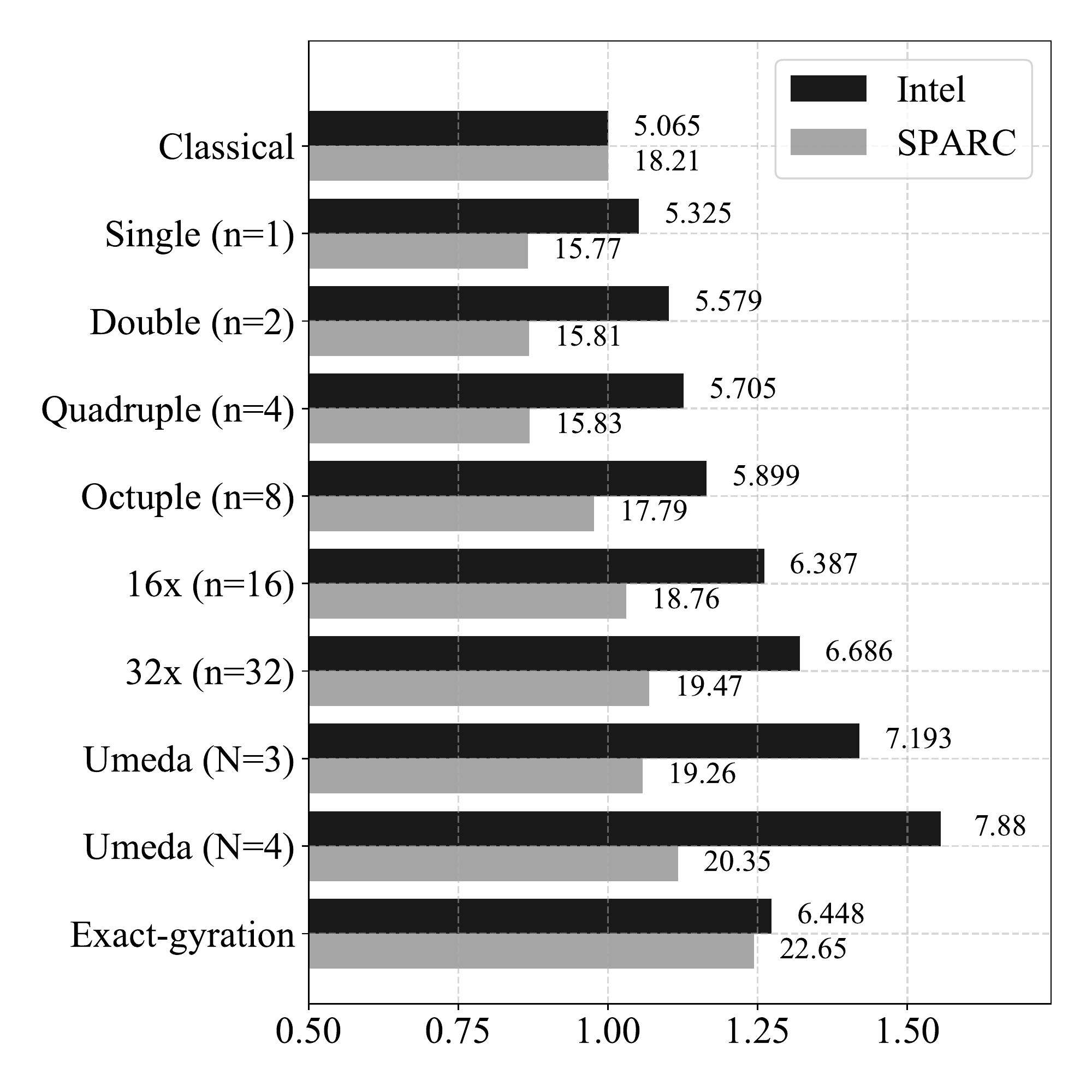}
\caption{
Benchmark results of the test-particle simulations.
The average elapse times (in seconds) are normalized to
those of the classical Boris solver.
\label{fig:bar}}
\end{center}
\end{figure}

In the second test,
we have measured elapse times of test-particle simulations. 
Since algorithms' performance relies on CPU architectures,
we have run the code on Intel and SPARC processors.
We have compiled the program on an Intel Core i7 processor with Intel Fortran compiler 2019.
For each solver, we have run the program for $1.2 \times 10^7$ timesteps. We have measured the elapse time six times for every cases.
The black bars in Figure \ref{fig:bar} show the results,
in normalized units.
We have also benchmarked the program on a SPARC processor with FUJITSU Fortran compiler on FX100 supercomputer at the Japan Aerospace Exploration Agency (JAXA). The elapsed times are averaged over three runs, because the results on the SPARC processor have less variations than those on the Intel processor. The results are presented in the gray bars in Figure \ref{fig:bar}.
To emphasize the difference, the leftmost value is set to 0.5.
Note that we show total elapse time of test-particle simulations.
The programs contain not only the Lorentz-force part, but also the Coulomb-force parts (Eqs.~\eqref{eq:first} and \eqref{eq:third}) and the position part (Eq.~\eqref{eq:x}).
Thus the results are not proportional to, but
highly affected by the computation time of the Lorentz-force solvers.

As far as we have tested, the new solvers are adequately fast.
On the Intel processor, they are slower than the classical Boris solver,
but faster than the Umeda solvers and the exact-gyration solver except for the $n=32$ case.
The multiple Boris solvers become slower as $n$ increases.
On the SPARC processor,
the new solvers are more promising than on the Intel processor.
Some of them ($n=1$--$4$) are 50\% faster than the exact-gyration solver and
even faster than the classical Boris solver.
It is noteworthy that
the double Boris solver ($n=2$) is significantly faster than the Umeda solver ($N=3$),
although the two solvers are equivalent. 
Similarly, the quadruple Boris solver ($n=4$) is faster than the Umeda solver ($N=4$).
Surprisingly, even the 32x Boris solver ($n=32$) looks as good as the Umeda solver ($N=3$),
even though it gives $256$ times higher accuracy.

\begin{table}
\begin{center}
\begin{tabular}{llll}
\hline
\#
~~
 & 
$\vec{B}$~~~~~~~~ &
$\vec{E}$~~~~~~~~ &
%$u_x$ &
~\\
\hline
1 &
(0,0,0) & (1,0,0) 	& {\rm direct acceleration by {\bf E}} \\
2 &
(0,0,0.1) & (1,0,0) 	& {\rm {\bf E}-dominated} \\
3 &
(0,0,1) & (1,0,0) 	& {\rm $|\vec{E}|=c|\vec{B}|$} \\
4 &
(0,0,1) & (0.1,0,0) & {\rm {\bf E}$\times${\bf B} drift} \\
5 &
(0,0,1) & (0,0,0) 	& {\rm gyration about {\bf B}} \\
\hline
\end{tabular}
\end{center}
\caption{
Field settings for test-particle simulations.
\label{table}}
\end{table}

\begin{figure}[htbp]
\begin{center}
\includegraphics[width={\columnwidth}]{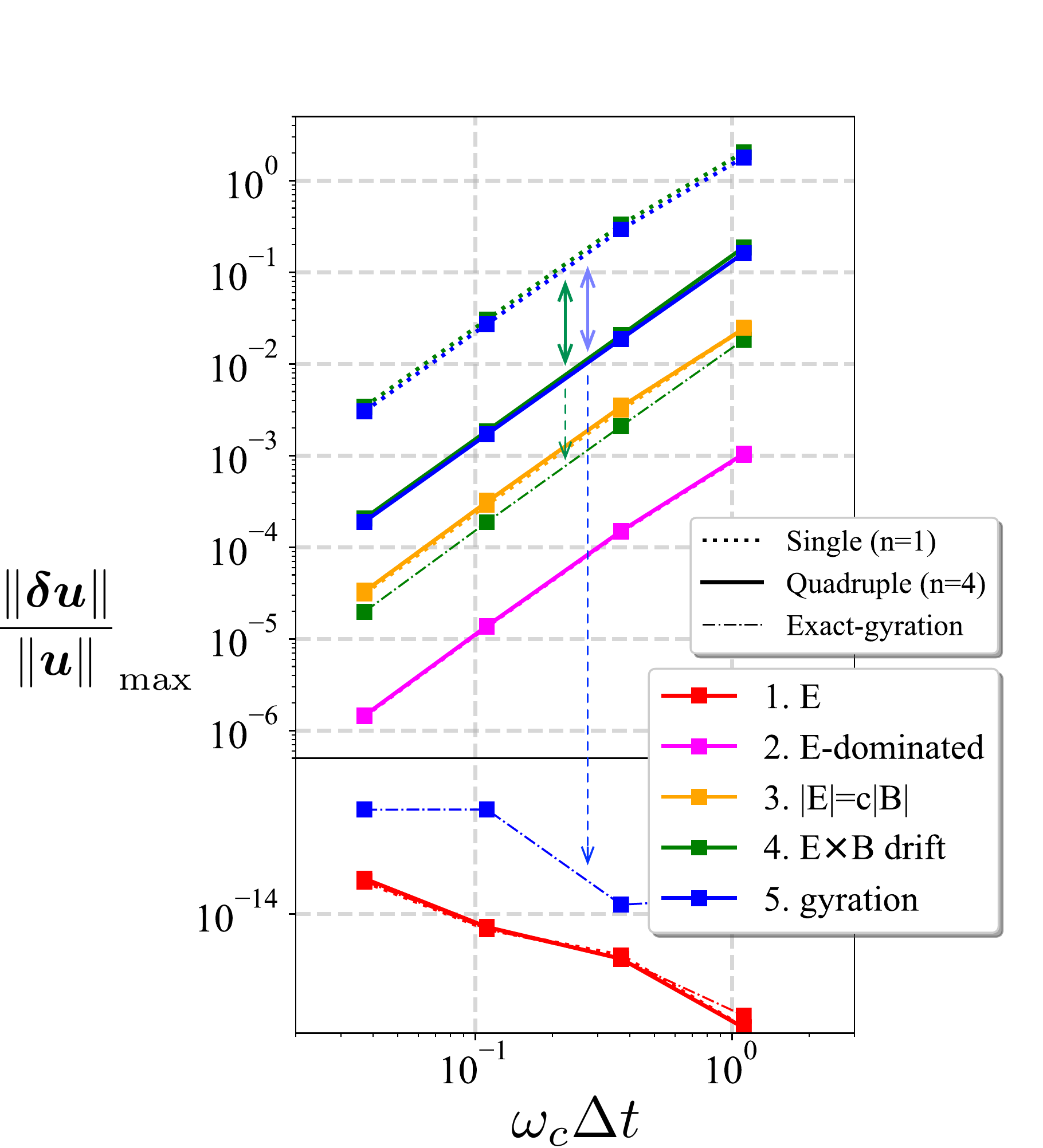}
\caption{
Normalized errors in $\vec{u}$ in the test-particle simulations, as a function of $\theta=\omega_c\Delta t$.
The single ($n=1$) Boris solver (dotted lines),
the quadruple ($n=4$) solver (solid lines) and
the exact-gyration solver (dashdotted lines)
are tested
in five configurations (colors; Table \ref{table}).
\label{fig:quadruple}}
\end{center}
\end{figure}

In the third test,
we have run test-particle simulations with
the single ($n=1$) Boris solver,
the quadruple ($n=4$) Boris solver, and
the exact-gyration solver in various configurations.
We consider five configurations, as shown in Table \ref{table}.
In cases 1 and 2, the electric field {\bf E} is dominant: $|\vec{E}| \gg c|\vec{B}|$.
In case 3, the electric and magnetic fields are equal: $|{E}|=c|{B}|$. 
In cases 4 and 5, the magnetic field {\bf B} is dominant: $|\vec{E}| \ll c|\vec{B}|$.
In particular, case 5 stands for the gyration about {\bf B}, as studied in the first test. 
Other parameters, timesteps, and the initial conditions
are similar to those in the first test.
We evaluate maximum values of the normalized error
$(| \delta\vec{u} | / | {u} |)_{\rm max}$
during the time interval of $0<t<12\pi$.
Here, $\delta\vec{u}$ is deviations from
analytic solutions (cases 1 and 5) or numerical reference values (cases 2--4).
We obtain the numerical reference values
by using the exact-gyration solver with $\Delta t = \pi/240$.
These configurations and procedures are
the same as in Ref.~\citep{zeni18b}.

Figure \ref{fig:quadruple} shows
the errors as a function of the timestep.
The dotted, solid, and dashdotted lines indicate
the results by the single ($n=1$) Boris solver, 
the quadruple ($n=4$) solver, and
the exact-gyration solver, respectively.
In case 1 (in red) by the three solvers and
case 5 by the exact-gyration solver,
one can essentially see no error. 
All the other curves suggest the second-order accuracy. 
In cases 2 and 3, the three solvers return similar results.
Meanwhile, in cases 4 and 5, their results are different.
The quadruple Boris solver provides
10.9--16.5 times smaller errors than the single Boris solver,
as emphasized by the solid arrows in green and blue. 
The improvement is asymptotic to $16$ for $\omega_c\Delta t \ll 1$ in case 5. 
The exact-gyration solver provides even better results,
as indicated by the dashed arrows. 
% [16.47625077 16.46982187 16.0886059  10.93558739] case 4
% [15.99691037 15.97181143 15.64265399 11.07011356] case 5
The quadruple Boris solver provides similar errors in cases 4 and 5.
So does the single Boris solver.
It appears that
the accuracy in the Lorentz-force solver (case 5) limits
the overall accuracy for $|\vec{E}| \ll c|\vec{B}|$ (case 4).

\begin{figure}[htbp]
\includegraphics[width={\columnwidth}]{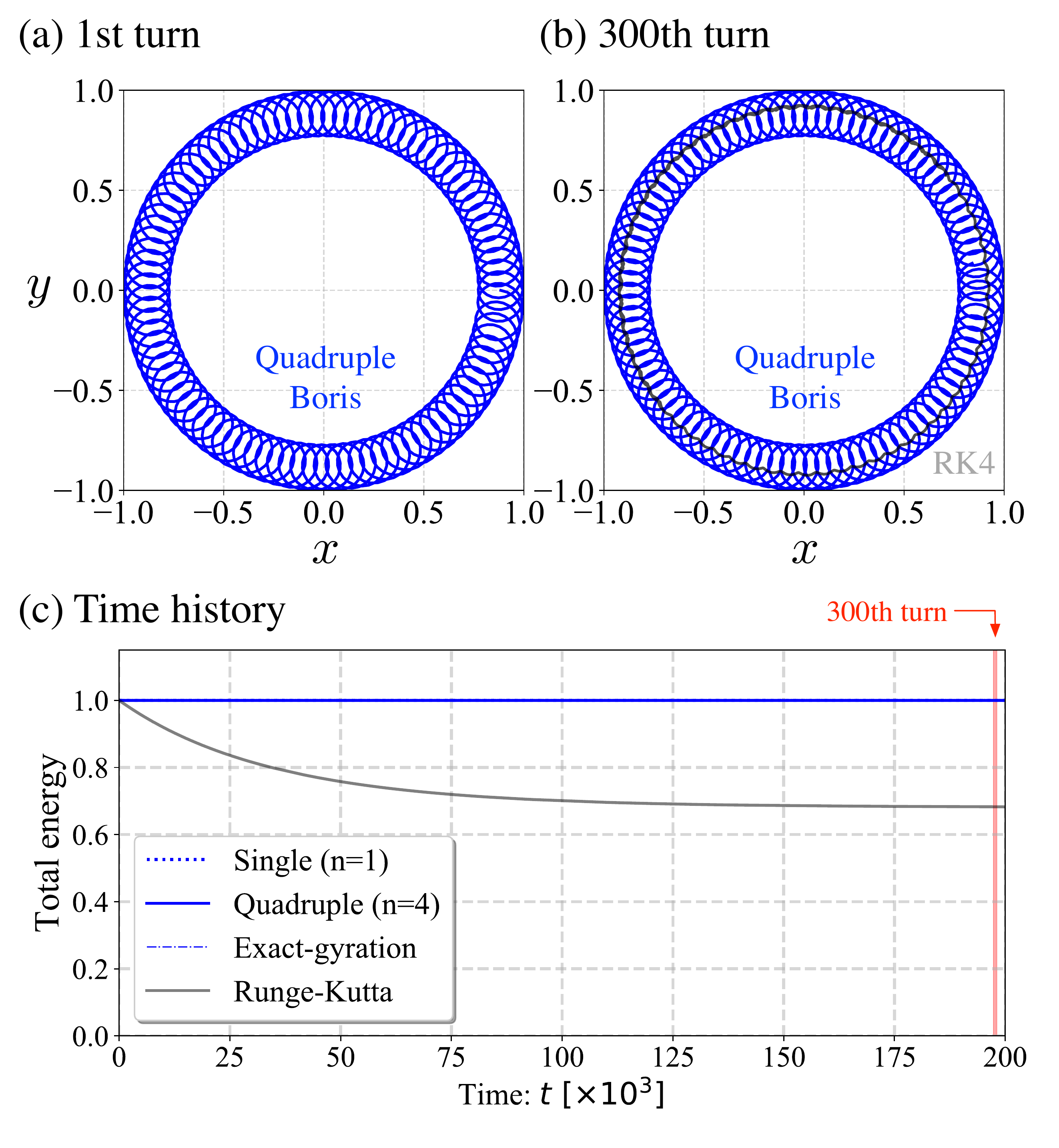}
\caption{
Particle trajectories by the quadruple Boris solver
at (a) the initial stage and (b) the late stage.
A trajectory by the RK4 solver is overplotted in gray at (b) the late stage.
(c) Energy histories by the four solvers.
\label{fig:qin}}
\end{figure}

In the fourth test, 
we study the long-term stability of the particle motion
in a non-uniform electromagnetic field. 
As done in Refs.~\cite{qin13,zeni18b},
we have run test-particles in the following field,
\begin{align}
\vec{B} = (x^2+y^2)^{1/2} \vec{e}_z,
~~
\phi = 0.01 (x^2+y^2)^{-1/2}
.
\end{align}
The conditions are set to $\vec{u}(t=0)=(0.1,0.0)$,
$\vec{x}(t=0)=(0.9,0,0)$, $m=c=1$, and $\Delta t = \pi/10$.
We compare the single ($n=1$) Boris solver, 
the quadruple ($n=4$) Boris solver,
the exact-gyration solver, and 
the fourth order Runge--Kutta solver.
The blue and gray lines in Figure \ref{fig:qin} show the trajectories
by the quadruple Boris solver and by the Runge--Kutta solver,
at (a) an initial stage and at (b) a late stage.
The time history of the total energy is presented in Figure~\ref{fig:qin}(c).
It combines the particle kinetic energy and the electric potential energy at the midpoint positions ($\vec{x}(t+\frac{\Delta t}{2}$)).
The values are normalized by the initial values.

One can see typical trajectories in Figure~\ref{fig:qin}(a).
It consists of a fast small-scale gyration and a slow large-scale rotation.
The large-scale rotation is due to the $\nabla B$ drift and the {\bf E}$\times{\bf B}$ drift. 
In the case of the quadruple Boris solver,
the particle keeps gyrating
even after a long time (300th turn; $t \approx 2\times 10^5$)
in Fig.~\ref{fig:qin}(b). 
The particle energy is excellently conserved.
We only recognize a relative error of $\mathcal{O}(10^{-4})$.
The trajectories (not shown), the time histories, and the error amplitudes (not shown) by the single ($n=1$) Boris solver and the exact-gyration solver are
similar to those of the quadruple ($n=4$) Boris solver.
In contrast, as evident in the gray line in Fig.~\ref{fig:qin}(b),
the particle gyroradius becomes smaller and smaller
in the case of the Runge--Kutta solver.
The particle numerically lost 30\% of its total energy (Fig.~\ref{fig:qin}(c)).

These results are attributed to
the volume-preserving property of the numerical schemes.
It is known that the Runge--Kutta solver is not volume preserving.
The solver is incapable of preserving
the energy of the small-scale oscillation.
On the other hand, as discussed in Section \ref{sec:multi},
the multiple Boris solvers are volume-preserving.
The other two solvers are also known to be volume-preserving \citep{qin13,zhang15,zeni18b}.
These three schemes are capable of preserving the energy of
the small-scale oscillation, even in a nonuniform field.

\section{Discussion and Summary}
\label{sec:discussion}

We have proposed the multiple Boris integrators
that virtually use the Boris procedure $n$ times in the Lorentz-force part.
The schemes provide second-order but
$n^{2}$-times smaller errors (Eq.~\eqref{eq:error})
than the classical Boris solver.
The coefficients are given by either
the generic forms of $p$ (Eqs.~\eqref{eq:p1}--\eqref{eq:p3}), the generic forms of $t$ (Eqs.~\eqref{eq:t1}--\eqref{eq:t3}), or the explicit forms (Eqs.~\eqref{eq:c1}--\eqref{eq:c4}). 
In our experience, the generic forms with $p$ are useful for large $n$.
For small $n$ ($n \le 4$), we prefer the explicit form,
because we can further optimize the code.
In our implementation for $n=1\dots 4$,
we have calculated the numerators and denominators in Eqs.~\eqref{eq:c1}--\eqref{eq:c4} separately,
because the explicit forms share the same denominator, $(1+t^2)^n$. 
As $\vec{t}$ contains $1/\gamma$ (Eq.~\eqref{eq:angle_n}),
we multiply the numerators and denominators by $\gamma^{2n}$.
By doing so, we reduce the number of division operations
to save the total computation cost.

The new solvers are developed in the same spirit as
in the multistep Boris solvers \citep{umeda18,umeda19},
which considered $2^{N-2}$-times smaller timesteps in the Lorentz-force part. 
In the multiple Boris solvers,
we virtually divide the rotation angle into $n$ Boris units.
This $n$ is arbitrary and not limited to $2^{N-2}$.
Thus one can choose the best $n$ for one's application.
In addition, we calculate the vector equation (Eq.~\eqref{eq:ut}) only once,
because we prepare the coefficients in advance.
The coefficients are rational fractions,
and so it is easy to calculate them.
Consequently, the program runs efficiently,
as evident in the benchmark results.

We have also evaluated numerical accuracy of several solvers
in general cases of $\vec{B}\ne 0$ and $\vec{E}\ne 0$. 
Figure \ref{fig:quadruple} demonstrates the second-order accuracy in most cases,
because the second-order Lorentz-force solvers are combined with
the Coulomb-force procedures (Eqs.~\eqref{eq:first} and \eqref{eq:third}) in an operator-splitting manner. 
In this study, since all the Lorentz-force solvers are combined with the same Coulomb-force procedures,
the numerical accuracy for $|\vec{E}| \ll c|\vec{B}|$ is essentially controlled by the Lorentz-force solvers,
even though the resulting schemes provide the same second-order accuracy. 
This is evident in cases 4 and 5 in Figure \ref{fig:quadruple} ---
the quadruple ($n=4$) Boris solver provides $\approx 16$ times better accuracy, and
the exact-gyration solver, which corresponds to the multiple Boris solver in the $n\rightarrow \infty$ limit,
provides even better accuracy. 
In general, the multiple Boris solvers have higher accuracy than the classic Boris solver,
and then they bring better accuracy in the $|\vec{E}| \ll c|\vec{B}|$ cases.
Importantly, we often consider $|\vec{E}| \ll c|\vec{B}|$ in practical applications in PIC simulation.

The new solvers may further reduce the computation cost,
by allowing larger timesteps.
In PIC simulation, the timestep $\Delta t$ is limited
by the plasma frequency $\omega_{p}$ and the gyrofrequency $\omega_{c}$.
The ratio of $\omega_{p}$ to $\omega_{c}$ depends on applications, and
both frequencies are limited to
$\omega \Delta t \lesssim \mathcal{O}(0.1)$ due to the numerical accuracy and
$\omega \Delta t < 2$ due to the numerical stability \citep{birdsall}.
If the gyrofrequency limits the timestep,
the multiple Boris solvers can relax the limitations for the accuracy. 
As evaluated in Eq.~\eqref{eq:deltat_1},
the $n$-tuple Boris solver allows $n$-times larger timestep,
because it repeats the classical Boris solver with a smaller timestep $(\Delta t/n)$ in the Lorentz-force part.
Actually, in the other parts of PIC simulation,
the timesteps are unchanged.
For example, we update the position (Eq.~\eqref{eq:x})
for $\Delta t$, not for $\Delta t/n$.
This fact makes the timestep condition more complicated.
The new condition is given by Eq.~\eqref{eq:deltat_2}.
This is more restrictive than Eq.~\eqref{eq:deltat_1}.

In the nonuniform electromagnetic field,
the multiple Boris solvers have a favorable property of the long-term stability.
Since they inherit a volume-preservation property from the Boris-type element, 
the new solvers are capable of preserving a small-scale oscillation,
as they should be. 
As far as we have tested, they conserve the total energy very well.
These facts give us further confidence to use the new solvers
in PIC simulations for practical applications.

Among the new solvers,
considering the accuracy, the timestep, and computation costs,
we recommend the quadruple ($n=4$) Boris solver with $\omega_c \Delta t \lesssim 0.5$
to the readers. 
One can develop higher-order solvers ($n \gg 32$)
to obtain even better accuracy.
In such a case,
we recommend the exact-gyration solver \citep{zeni18b} to the readers. 
As shown in Figure \ref{fig:bar},
the 32x Boris solver ($n = 32$) is already more expensive than
the exact-gyration solver on Intel processors.
It is likely that higher-order solvers are
as expensive as the exact-gyration solver on SPARC processors for $n \gg 32$.
In any case, since the extra cost is limited,
the exact-gyration solver will be another choice for $n > 10$.  
Finally, we note that
our recommendation is partly based on
the benchmarks on our specific systems.
Performance may be different on other systems
such as GPU-based systems and FGPA systems.
In general, since it is impossible to predict future computing platforms,
we can only prepare as many particle solvers as possible.
Our solvers will be the latest additions.

In summary, we have proposed
the multiple Boris solvers that
virtually combine multiple Boris units in the Lorentz-force part.
We have derived one-step expressions
with polynomial-based coefficients
for arbitrary $n$ divisions.
The new solvers provide second-order,
$n^2$-times smaller errors than
the classical Boris solver.
They also allow larger timesteps.
Their performance is promising.
They are stable over a long time.
We hope that the proposed solvers will be useful in
studies with PIC simulations.

\section*{Acknowledgements}
This work was partially carried out on facilities of the JSS2 system at Japan Aerospace Exploration Agency (JAXA). This work was supported by
Grant-in-Aid for Scientific Research (C) 17K05673, (S) 17H06140, and (B) 17H02877
from the Japan Society for the Promotion of Science (JSPS).

\clearpage
\appendix

\section{Third coefficients for $n=2k+1$}
\label{sec:2k+1}

We recall the following relation of trigonometric functions.
\begin{align}
\sin( A + \alpha ) +
\sin( A - \alpha )
=
2 \sin A \cos \alpha
\end{align}
We substitute $(2k+1)\alpha$ to $A$,
where $k$ is a positive integer, $k\ge 1$.
Using Chebyshev polynomials, we obtain
\begin{align}
2\sin\big((2k+1) \alpha\big)
\cos\alpha
&=
\sin\big(2(k+1) \alpha\big)
+
\sin\big(2k \alpha\big)
\nonumber \\
&=
\sin 2\alpha
\Big(
U_{k}(\cos 2\alpha)
+
U_{k-1}(\cos 2\alpha)
\Big)
\\
\sin\big((2k+1) \alpha\big)
&=
\sin \alpha
\Big(
U_{k}(\cos 2\alpha)
+
U_{k-1}(\cos 2\alpha)
\Big)
\end{align}
With help from Eqs.~\eqref{eq:trig1}, \eqref{eq:trig2}, and \eqref{eq:p},
we obtain
\begin{align}
c_{n3}
&=
\frac{2}{t^2}
\sin^2\big((2k+1) \alpha\big)
\nonumber \\
&=
\frac{2}{1+t^2}
\Big\{
U_{k}\Big(\frac{1-t^2}{1+t^2}\Big)
+
U_{k-1}\Big(\frac{1-t^2}{1+t^2}\Big)
\Big\}^2
\nonumber \\
&=
(1+p)
\Big\{
U_{k}(p)
+
U_{k-1}(p)
\Big\}^2
\end{align}

\end{document}